# Modeling Aspect Mechanisms: A Top-Down Approach[*]

Sergei Kojarski[†]     David H. Lorenz
Department of Computer Science, University of Virginia
Charlottesville, Virginia 22904-4740, USA

{kojarski,lorenz}@cs.virginia.edu

## ABSTRACT

A plethora of diverse aspect mechanisms exist today, all of which integrate concerns into artifacts that exhibit crosscutting structure. What we lack and need is a characterization of the design space that these aspect mechanisms inhabit and a model description of their weaving processes. A good design space representation provides a common framework for understanding and evaluating existing mechanisms. A well-understood model of the weaving process can guide the implementor of new aspect mechanisms. It can guide the designer when mechanisms implementing new kinds of weaving are needed. It can also help teach aspect-oriented programming (AOP). In this paper we present and evaluate such a model of the design space for aspect mechanisms and their weaving processes. We model weaving, at an abstract level, as a concern integration process. We derive a weaving process model (WPM) top-down, differentiating a reactive from a nonreactive process. The model provides an in-depth explanation of the key subprocesses used by existing aspect mechanisms.

## Categories and Subject Descriptors

D.2.10 [**Software Engineering**]: Design; D.1.5 [**Programming Techniques**]: Aspect-oriented Programming; D.3.2 [**Programming Languages**]: Language Classifications

## General Terms

Design, Languages

## Keywords

AOP, AspectJ, DFD, Hyper/J, Open Classes, aspect mechanism, crosscutting concerns, definition, nonreactive, reactive, taxonomy, top-down classification, weaving process model (WPM).

[*]This research was supported in part by NSF's Science of Design program under Grants Number CCF-0438971 and CCF-0609612.
[†]Sergei Kojarski is a PhD candidate at Northeastern University and a visiting graduate student at University of Virginia.

## 1. INTRODUCTION

Concern integration is an essential process in aspect-oriented programming (AOP). Kiczales *et al.* [6] explicitly express the goal of AOP in terms of combining separate crosscutting concerns:

> *"The goal of Aspect-Oriented Programming is to make it possible to deal with cross-cutting aspects of a system's behavior as separately as possible. We want to allow programmers to first express each of a system's aspects of concern in a separate and natural form, and then automatically combine those separate descriptions into a final executable form using a tool called an Aspect Weaver."*

The process of concern integration is called *weaving*, and the implementation of an "aspect weaver"—an *aspect mechanism*. This paper presents a top-down model of an aspect mechanism and its internal weaving process.

### 1.1 Evolution of AOP Models

A model primarily reflects understanding. We identify the following evolutionary stages in the common understanding of AOP:

1. ($\mathbf{A} \times \mathbf{B} \to \mathbf{B}$) In early AOP models, an aspect mechanism was explained in terms of compilation semantics. This led to thinking about AOP in terms of code instrumentation, where an aspect program ($p_A \in \mathbf{A}$) is woven into a base program ($p_B, p'_B \in \mathbf{B}$). We refer to this initial understanding of AOP as the ABB model (Figure 1).

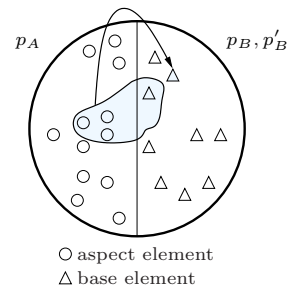

**Figure 1: The ABB Model**

2. ($\mathbf{A} \times \mathbf{B} \to \mathbf{X}$) Later, code instrumentation was found to be misleading and confusing in explaining the essence of AOP. When AOP was better understood, the compilation semantics was replaced with dynamic weaving semantics. The main insight was that crosscutting is an emerging property found in the composed program ($p_X \in \mathbf{X}$) rather than a pre-existing property in either $p_A \in \mathbf{A}$ or $p_B \in \mathbf{B}$. We refer to this observation as the ABX model (Figure 2), named to correspond with Masuhara and Kiczales' ABX model of crosscutting [14].[1]

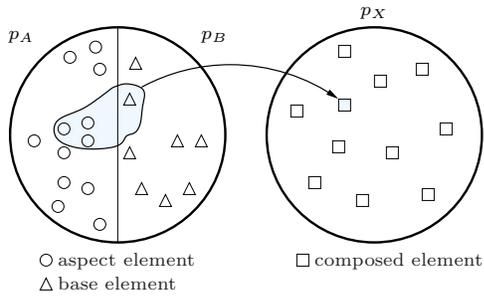

**Figure 2:** The ABX Model

3. ($\mathbf{C} \times \mathbf{R} \to \mathbf{X}$) More recently, the syntactical distinction between aspect ($\mathbf{A}$) and base ($\mathbf{B}$) has begun to diminish. In modern AOP languages, such as AspectWerkz [1] and Classpects [17], there is no essential distinction between aspects and classes (as presumably will also be the case in AspectJ 5). The traditional, syntactical dichotomy into $p_A \in \mathbf{A}$ and $p_B \in \mathbf{B}$ is being slowly replaced with a semantical distinction between concerns $p_C \in \mathbf{C} = \mathbf{A} \cup \mathbf{B} \setminus \mathbf{R}$, expressed in either $\mathbf{A}$ or $\mathbf{B}$, and integration rules $p_R \in \mathbf{R}$, expressed in code, annotations, XML, or some other form. We introduce and refer to this model as CRX (Figure 3).

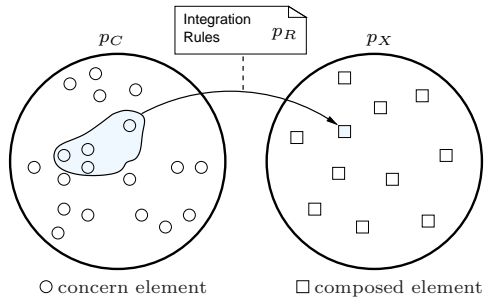

**Figure 3:** The CRX Model

Interestingly, the CRX model is very close to the original Hyper/J model [16], taking this evolutionary path a full circle. The introduction of the CRX model as a general AOP model that explains both Hyper/J-like and AspectJ-like aspect mechanisms is a contribution of this paper and a direct result of a top-down approach.

## 1.2 Problems with Current Models

Models of aspect mechanisms have focused either on a selected mechanism, or on a bottom-up generalization of several mechanisms. In a bottom-up abstraction, a model is built by identifying and generalizing common patterns. Since each aspect mechanism is useful precisely because it does some crosscutting thing very different, abstracting over several aspect mechanisms is difficult and often results in a fine-grained model which would likely need to be further extended to fit future aspect mechanisms.[2]

We distinguish between two general approaches to modeling aspect mechanisms, namely, semantical and conceptual (Table 1, two left columns). Existing *semantical models* either explain a specific AspectJ-like Pointcut and Advice (PA) mechanism or generalize a PA model. Lämmel [8] explains a join-point-and-advice mechanism named Method-Call Interception (MCI). Wand *et al.* [22, 15] explain an aspect mechanism for AspectJ. Walker *et al.* [21] define aspects through explicitly labeled program points and first-class dynamic advice. Jagadeesan *et al.* [5] use PA to define AOP functionality. Although very precise, all of these semantical models do not generalize over non-PA aspect mechanisms.

An example of a *conceptual model* is presented by the work of Masuhara and Kiczales [14] on modeling crosscutting in aspect-oriented mechanisms. Their *aspect sand-box* (ASB) [14, 20] framework generalizes over four mechanisms: Pointcut and Advice (PA), Open Classes (OC), Traversal (TRV), and Compositor (CMP). PA and OC are implemented by AspectJ; TRV is found in Demeter; and CMP is realized by Hyper/J.

ASB belongs to the category of ABX models. Generally, ASB represents each aspect mechanism as a *weaver* that combines an aspect program and a base program into a result computation (or program) $X$. In ASB, a mechanism realizes a function with the signature:

$$A \times B \times METAinteger \to X$$

where $METAinteger$ stands for composition rules existing only in CMP.[3]

The weaver is said to model a *weaving process*, which is defined informally as [14]:

> "*taking two programs and coordinating their coming together into a single combined computation.*"

The weaver's semantics is presented by an 11-tuple structure:

$$\langle X, X_{\text{JP}}, A, A_{\text{ID}}, A_{\text{EFF}}, A_{\text{MOD}}, B, B_{\text{ID}}, B_{\text{EFF}}, B_{\text{MOD}}, METAinteger \rangle$$

where $A$ and $B$ denote the languages of the input programs, $X$ is the result domain of the weaving process, and $X_{\text{JP}}$ are join points in $X$. The elements $A_{\text{ID}}, A_{\text{EFF}}$ of $A$ and $B_{\text{ID}}, B_{\text{EFF}}$ of $B$ provide a common frame of reference. Programs in $A$ and $B$ refer to $X_{\text{JP}}$ using $A_{\text{ID}}$ and $B_{\text{ID}}$ and contribute their $A_{\text{EFF}}$ and $B_{\text{EFF}}$ effects, respectively, to the semantics of the corresponding $X$ computations (program declarations). The remaining two elements, $A_{\text{MOD}}$ and $B_{\text{MOD}}$, refer to structures of modularity in the input languages.

The introduction of a result domain $X$ (that also coincides with our CRX model) is the main contribution of ASB. In ASB, join points exist only in $X$. An aspect mechanism combines semantics contributed by $A$ and $B$ at join points in $X$ using the common frame of reference. $A$ and $B$ do not crosscut each other directly, but only with respect to $X$.

The essential shortcomings of ASB are:

- **Syntactical dichotomy.** ASB defines aspect mechanisms over an AOP language of a fixed *syntactical* form. Specifically, an AOP program is required to consist of an aspect program $p_A \in A$ and base program $p_B \in B$, both specifying concerns. When the syntax of the AOP language is different, the model includes it as a special case. Most noticeable, the *META* component was added so that ASB can also model Hyper/J.

---

[1] For consistency with the other models and to avoid confusion we use $\mathbf{A}$ to denote the *aspect* language and $\mathbf{B}$ to denote the *base* language, whereas Masuhara and Kiczales originally used $\mathbf{A}$ for the base language and $\mathbf{B}$ for the aspect language.

[2] This is a criticism of the term *aspect-oriented* being used today to describe a broad array of diverse programming mechanisms, rather than a criticism of bottom-up generalization per se.

[3] For all of the other mechanisms, $METAinteger$ is left out of the model.

|  | Other Models | | Our Model |
|---|---|---|---|
| **Approach** | **Semantical** | **Conceptual** | **Software Engineering** |
| **Design level** | Implementation | Analysis | Analysis and Design |
| **Generalization method** | - | Bottom-Up | Top-Down |
| **Representative example** | MCI [8] | ASB [14, 20] | CRX |
| **Mechanisms modeled** | PA | $PA_{AJ}$, OC, CMP, TRV | PA, OC, CMP |
| **Unifying concept** | - | Join points, JPM | Integration rules, WPM |
| **Dichotomy** | Semantical | Syntactical | Semantical |
| **Organizational structure** | - | Flat | Hierarchical |
| **Resolution** | Fine-grained | Coarse-grained | Adjustable |
| **Abstract** | No | Yes | Yes |
| **Precise** | Yes | No | Yes |
| **Uniform** | - | No | Yes |

Table 1: **Comparison of approaches to modeling aspect mechanisms**

The definition of an aspect mechanism in ASB does not generalize neatly over languages with similar semantics but a different syntactical structure. For example, Classpects [17] has no aspects, just classes. AspectWerkz [1] has no advice, just methods. In AspectWerkz, a method annotated as advice can be invoked explicitly as a method or implicitly as advice. This duality is difficult to explain in an ABX model due to its firm syntactical distinction between aspect ($p_A \in A$) and base ($p_B \in B$) programs. Furthermore, the annotations in AspectWerkz can be extracted and placed in a separate XML file. These XML integration rules are not in $A$, $B$, or $X$.

- **AspectJ-based abstraction.** ASB selects $X_{JP}$ as the central abstraction found in all aspect mechanisms. In ASB terms, the common frame of reference is $X_{JP}$ and the programs connect at the join points. While the concept of a join point is essential in PA, it is found in only a subset of existing aspect mechanisms. Specifically, OC may be explained with or without join points (Section 5); it is unclear that join points are the right abstraction for TRV [9]; and CMP does not involve join points in $X$ at all [14, 20].

- **Coarse-grained process.** In ASB, a join point in $X_{JP}$ describes a computation in $X$, and the computation in $X$ is constructed using the join point in $X_{JP}$. Obviously, an aspect mechanism must construct the join point *prior* to the construction of the corresponding computation it describes. ASB does not explain how this cyclic dependency is resolved, and generally lacks an explicit weaving process model.

- **Over generalization.** ASB takes upon itself to abstract over four mechanisms that do not necessarily share properties that can be reasonably generalized. PA and CMP are *oblivious* [4] and provide *integration* mechanisms. TRV, on the other hand, is *not* oblivious and provides an *adaptation* mechanism [9]. Yet, ASB generalizes over all of them [20]. To reconcile the difference between these four mechanisms, ASB provides a fine-enough grained structure. ASB's bottom-up generalization underscores the similarities between various aspect mechanisms. What is lost in the generalization, however, is the ability to also understand their *differences*.

## 1.3 Contribution

This paper presents an abstraction in the opposite direction. Using a *top-down* approach, we derive an abstract, yet precise, architectural model of the design space for aspect mechanisms. The model is build through analysis and design. The analysis phase characterizes the external behavior of an aspect mechanism. The design phase models the mechanism's internal behavior.

The novelty of our model lies in its *semantical* rather than syntactical generalization. We define the aspect mechanism's domains over a semantical representation of an AOP program, which can be constructed from various syntactical forms. Our model explains the essential differences between various aspect mechanisms. It categorizes aspect mechanisms according to semantical operations.

Our CRX model emphasizes integration. The integration logic is defined by integration rules, separate from concern code defined by aspect or base programs. The CRX model provides a natural abstraction for integration-based weaving processes. It captures the similarities as well as the differences between aspect mechanisms on various levels of abstraction. The top-down specialization respects individual features of different aspect mechanisms. For example, at the top level, the Hyper/J and AspectJ mechanisms are similar; and a deeper level reveals how different they are.

## 1.4 Outline

In Section 2 we analyze the weaving problem and present a general CRX weaving process model (WPM). The problem and the model are defined in abstract terms. They are constructed top-down, independent from any particular aspect mechanism. The top-down construction introduces different categories of a weaving process. In Sections 3 and 4 we design a nonreactive and a reactive specialization of CRX, respectively. We use concrete aspect mechanisms to exemplify each category. In Section 5 we explain the duality of AspectJ's OC with respect to these categories. In Section 6 we compare the CRX model to the ABX model, in terms of providing a deeper explanation of AOP. Section 7 discusses how our classification of weaving processes is different from existing categorizations of AOP languages. Finally, Section 8 discusses how the model can help in constructing new aspect mechanisms.

## 2. ANALYSIS

This section explains the fundamentals of the concern integration process at an abstract level. We begin by articulating what a *weaving problem* is, and what is a solution, termed a *weaving plan*, to a weaving problem. We then present a high-level model of a *weaving process* that explains the external behavior of an aspect mechanism.

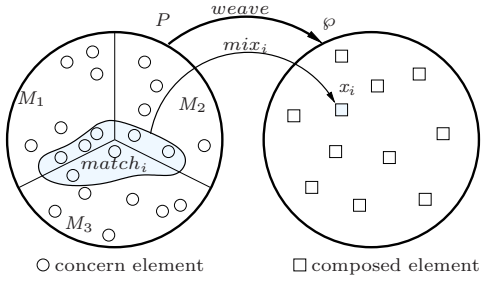

Figure 4: The Weaving Problem

## 2.1 Weaving Problem

An aspect mechanism implements a solution to a problem of concern integration. Intuitively, before separate concern elements can be "computed" they need to be woven together to produce a "computable" composed element. That is the job of an aspect mechanism.

Per program, the problem of concern integration is to map the modularized program $P$, in which concerns are separated, to a composed program $\wp$, in which the concerns are woven (Figure 4). We denote the mapping by:

$$P \stackrel{weave}{\longmapsto} \wp$$

Abstractly, a *program* is either *code* or *computation*. A concern program,

$$P = \{c_1, \ldots, c_n\}$$

is a set of $n$ program elements, $c_j \in \mathcal{C}$, $j = 1, \ldots, n$, where $\mathcal{C}$ is a domain of concern elements. A modularized concern program $P$ is partitioned into pairwise-disjoint concern modules. For example, in Figure 4, $P = M_1 \cup M_2 \cup M_3$.

A composed program,

$$\wp = \{x_1, \ldots, x_m\}$$

is a set of $m$ program elements, $x_i \in \mathcal{X}$, $i = 1, \ldots, m$, where $\mathcal{X}$ is a domain of composed elements and where the crosscutting occurs.

## 2.2 Weaving Plan

A solution to a concern integration problem is a *weaving plan*. A weaving plan establishes a mapping between elements of the concern program $P$ and elements of the composed program $\wp$. A weaving plan is specified by a set of integration rules,

$$\varrho = \{r_1, \ldots, r_k\}$$

where $\mathcal{R}$ is a domain of integration rules, and $r_i \in \mathcal{R}, i = 1, \ldots, k$. The meaning of a weaving plan is a set of pairs,

$$[\![\varrho]\!] = \{\langle match_i, mix_i \rangle : x_i \in \wp\}$$

where for every element $x_i$ in $\wp$, $match_i \in 2^P$ is a subset of program elements, $mix_i : 2^P \hookrightarrow \wp$ is a constructor, and the weaving plan satisfies the condition that $x_i \in \wp$ is constructed by applying $mix_i$ to $match_i$. We denote this mapping by:

$$match_i \stackrel{mix_i}{\longmapsto} x_i$$

## 2.3 Weaving Process

We model weaving as a concern integration process. An aspect mechanism constructs $\wp$ by integrating concerns in $P$ according to a plan $\varrho$. In our model, the aspect mechanism implements a CRX weaving process (Figure 5). At each step of the process, the mechanism accesses elements of $P$ (the $\mathcal{C}$-register), elements of $\varrho$ (the $\mathcal{R}$-register), and elements of the about-to-be-composed program $\wp$ (the $\mathcal{X}$-register). The mechanism produces new elements of the composed program $\wp$, and writes them to the $\mathcal{X}$-register.

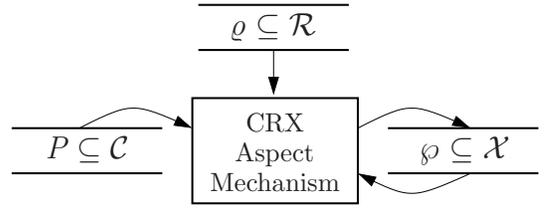

Figure 5: CRX Weaving Process

A weaving process is said to be *nonreactive* (Section 3) if does not depend on the state of the $\mathcal{X}$-register. A weaving process is said to be *reactive* (Section 4) if it requires read-access to the $\mathcal{X}$-register in order to produce the composed elements.

The top level CRX process is the most general model of an aspect mechanism. In the next two sections we describe two specializations of weaving process. We explain the semantics of these models by mapping them to well known aspect mechanisms.

## 2.4 Notation

We describe the weaving process using a sequence of data flow diagrams (DFDs) [24, 18, 23]. A DFD is a graph that models the data flow communication among processes and data stores. Circles represent transformations. Pairs of parallel lines represent data stores. Arrows represent data. Collectively, a sequence of DFDs specifies the behavior of an aspect mechanism in terms of its internal processes that transform data and the type of data being transformed. At the top level, a context diagram describes the overall system and its data flow interfaces with its environment. The context diagram is then decomposed into level-1 processes, which are then decomposed into level-2 processes, and so on.

We use the standard DFD convention for labeling processes. A context level DFD is numbered $n.0$. Level-1 processes are labeled $n.1$, $n.2$, etc. Successive nested levels follow a similar numbering convention. This convention provides for easy tractability. We use the leading number $n$ to relate a sequence of diagrams to the section where they are discussed. The first time a process is mentioned in the text, we indicate its label in parentheses.

Note that a DFD does not specify when communications take place. In particular the labels are just identification tags and do not imply any specific processing order.

## 3. DESIGN OF A NONREACTIVE WPM

At each step of the weaving process, a nonreactive aspect mechanism computes an element (or several elements) of the composed program $\wp$ only from elements of $P$ and $\varrho$. In other words, the weaving plan of a nonreactive aspect mechanism does not consult the state of the $\mathcal{X}$-register in order to specify composed elements. In this section we informally introduce the inner working of a nonreactive aspect mechanism through an overview of the weaving process in Hyper/J.

### 3.1 A Nonreactive Aspect Mechanism

The CMP mechanism [16] of Hyper/J [19] is essentially a Java program transformer. In Hyper/J's terms, the input program $P$ is called a *hyperspace*, the concerns are called *hyperslices* (a hyperspace is a set of hyperslices corresponding to, e.g., $M_1$, $M_2$, $M_3$ in Figure 4), and the composed program $\wp$ is called a *hypermodule*. The composition logic, called *composition rules*, is the weaving plan $\varrho$ (Figure 6). $P$ keeps the concerns separate, but does not specify a coherent behavior. CMP generates $\wp$ with the desired behavior by integrating the concerns together according to the composition rules $\varrho$.

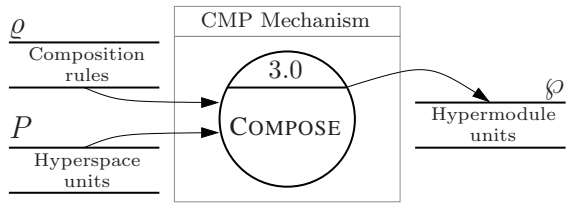

**Figure 6: Nonreactive CMP mechanism**

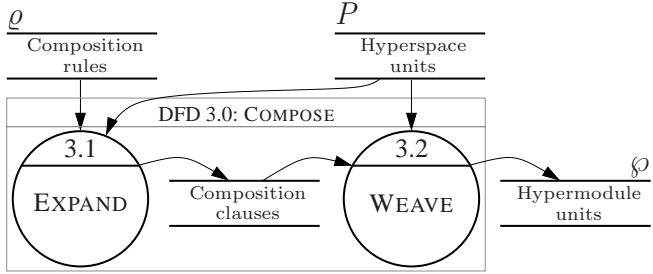

**Figure 7: DFD 3.0: COMPOSE**

**Listing 1: Personal View**
```
package personal;
public class Person {
 protected String name;
 protected String dob;
 public String getName(){return name;}
 public void setName(String name){this.name=name;}
 public String getDoB(){return dob;}
 public void setDOB(String dob){this.dob=dob;}
}
```

**Listing 2: Tax View**
```
package taxes;
public class Person {
 protected String name;
 protected String ssn;
 public String getName(){return name;}
 public void setName(String name){this.name=name;}
 public String getSSN(){return ssn;}
 public void setSSN(String ssn){this.ssn=ssn;}
}
```

**Listing 3: Hyperspace Specification**
```
–hyperspace
 hyperspace Person
   composable class personal.*;
   composable class taxes.*;
–concerns
 package personal: PersonalView
 package taxes: TaxView
```

**Listing 4: Composition Rules**
```
hypermodule Result
 hyperslices: PersonalView, TaxView;
 relationships: mergeByName;
end hypermodule;
```

**Listing 5: Composition Clauses**
```
hypermodule Result

operations
 getName: equivalent(signatures(<PersonalView.getName,
     TaxView.getName>))
 getDoB: identity(signatures(<PersonalView.getDoB>))
 getSSN: identity(signatures(<TaxView.getSSN>))
 //the remainder of the operations code is omitted

classes
 class Person
  instance variables:
    name: equivalent(types(<PersonalView.Person.name,
       TaxView.Person.name>))
    dob: identity(types(<PersonalView.Person.dob>))
    ssn: identity(types(<TaxView.Person.ssn>))

mapping
 getName: class Person
   CallAction: Sequence
     PersonalView.getName.Person
     TaxView.getName.Person
 getDoB: class Person
   CallAction: Simple PersonalView.getDoB.Person
 getSSN: class Person
   CallAction: Simple TaxView.getSSN.Person
 //the remainder of the mapping code is omitted
```

The composition rules are specified in terms of *unit trees*. A unit tree is an abstraction of a program's abstract syntax tree (AST), where units represent nodes. In general, the rules define each hypermodule unit as a composition of multiple hyperspace units.

We explain the COMPOSE (3.0) process through a coding example. Consider two classes describing a person, one from a personal perspective (Listing 1); another from a tax perspective (Listing 2). We can use Hyper/J to produce an integrated view. We start by specifying a hyperspace and a mapping of the classes to hyperslices (hyperspace concerns, Listing 3). The Person hyperspace defines two hyperslices, namely, PersonalView and TaxView. The former consists of units found in the personal Java package, and the latter comprises units of the taxes package.

The composition rules (Listing 4) define the Result hypermodule as a composition of the PersonalView and TaxView hyperslices under the **mergeByName** composition strategy. The strategy specifies that same-name-same-type hyperspace units (e.g., the class units PersonalView.Person and TaxView.Person; or the method units PersonalView.Person.getName and TaxView.-Person.getName) should be merged in the hypermodule.

Generally, the meaning of a composition rule depends on the structure of the input hyperspace. In our example, the result of the composition rules depends entirely on the hyperspace structure.

The semantics for Hyper/J's composition rules [16] define a two-

step composition process. The DFD in Figure 7 is an explosion of
COMPOSE (3.0). The first step is realized by the EXPAND (3.1) process, which expands the composition rules against the hyperspace
unit tree into *composition clauses* (Listing 5).[4] The composition
clauses specify a hypermodule as a composition of specific hyperspace units. At the second step, the composition clauses and the
hyperspace units are passed to the WEAVE (3.2) process, which
composes the result hypermodule.

In Listing 5, the operations section specifies how to compose the
hypermodule operations (method signatures) from signatures of the
hyperspace methods. The classes section defines a hypermodule's
class-graph (classes and their instance variables) as a composition
of hyperspace class-graph nodes. The mapping section introduces
methods into the hypermodule classes by associating a hypermodule operation with a set of hyperspace realizations (method-body
expressions) and a class. Thus, a composition clause defines a hypermodule unit as a combination of specific hyperspace units.

The weaving process is driven by the composition clauses. At
each step ($i$), WEAVE (3.2) selects and applies a clause. The composition clause specifies what hyperspace units to combine ($match_i$)
and how to combine them ($mix_i$). The process completes the step
by composing a new hypermodule unit (composed element $x_i$) and
committing it to the composed program. The process terminates
after all clauses were applied.

## 4. DESIGN OF A REACTIVE WPM

In this section we model a reactive CRX process. We present the
model through an explanation of AspectJ's PA mechanism.

### 4.1 A Reactive Aspect Mechanism

The context level DFD in Figure 8 depicts a reactive weaving
process of a PA mechanism. PA realizes a program execution process. Given a concern program $P$, a weaving plan $\varrho$, and a computation trace $\wp$, WEAVE (4.0) executes the program by constructing,
transforming, and running computations.

$P$ and $\varrho$ are constructed from an input program string. $P$ consists
of all the methods and advice-body expressions found in either the
base (e.g., Java classes) or the aspect (e.g., AspectJ aspects) parts
of the input program. $\varrho$ is constructed from three sources:

(1) pointcut designators;

(2) pointcut-to-advice mapping; and

(3) advice-to-type mapping.

The first two elements provide collaboratively a mapping from join
points to advice-body expressions. The last element provides a
mapping from an advice-body expression to its advice type (**before**,
**after**, or **around**), which is necessary for a proper weaving.

$\wp$ is a vector of computations that constitutes the composed program. WEAVE executes the program $\wp$ by constructing computations and running them. At each step of the process, WEAVE reads
in the most recent computation in the trace, and executes it. The
execution produces either a null-computation that wraps around a
value or a set of sub-computations that extend $\wp$.

Figure 9 is an explosion of WEAVE (4.0). ADVISE (4.1) and
EVALUATE (4.2) are level-1 subprocesses. EVALUATE realizes an
*expression evaluator* (a Java interpreter). This process executes the
most recent $\wp$ computation, and generally produces one or more
*join point* computations $x_{jp}$. The ADVISE process intercepts a join
point computation $x_{jp}$, and transforms it by wrapping it with advice

---
[4]For readability, the actual composition clauses produced by the
Hyper/J compiler were amended to resemble the notation in [16].

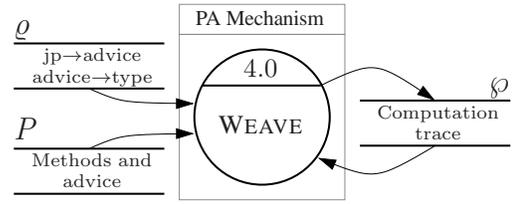

**Figure 8: Reactive PA Mechanism**

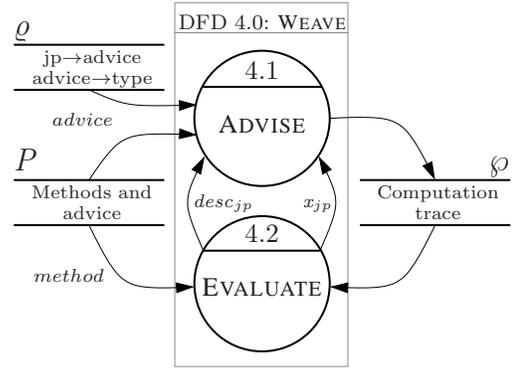

**Figure 9: DFD 4.0:** WEAVE

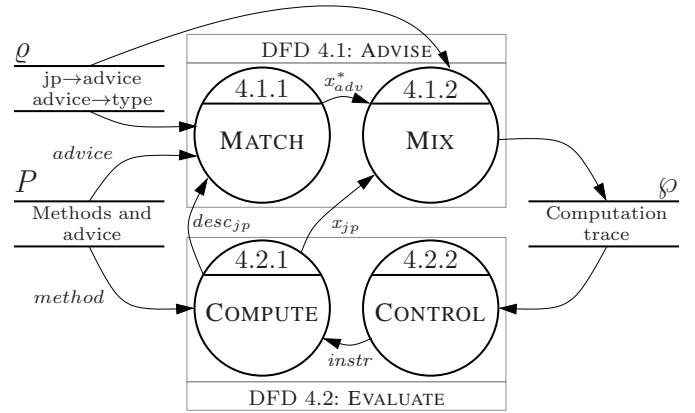

**Figure 10: DFD 4.1:** ADVISE **and DFD 4.2:** EVALUATE

computations. The computation produced by the ADVISE process
replaces $x_{jp}$ in the program execution.

The PA mechanism represents a join point computation $x_{jp}$ as
a *join point description* $desc_{jp}$. The description provides the type
of the join point computation (e.g., method call, method execution,
constructor execution), values drawn from its evaluation context,
and related static and lexical data. The join point description is a
means by which EVALUATE passes information about the program
state to ADVISE. The ADVISE process then uses this information
for selecting the appropriate advice.

A further breakdown of the PA's weaving process is depicted in
Figure 10. Four subprocesses constitute a reactive weaving process. MATCH (4.1.1) and MIX (4.1.2) realize advice selection and
advice weaving, respectively. MATCH selects pieces of advice by
applying the jp→advice mapping to $desc_{jp}$. The process returns the selected advice-body expressions as advice computations $x^*_{adv}$. MIX combines the advice computations with respect to their types with a join point computation $x_{jp}$ obtained from COMPUTE (4.2.1). The computation composed by MIX extends the execution trace $\wp$.

The CONTROL (4.2.2) process runs the most recent computation in $\wp$. In general, the computation is built from a complex expression. Running the computation requires evaluation of the expression's subterms. CONTROL represents subterms as instructions, and delegates their evaluation to COMPUTE. Since the process runs MIX-computations, the instruction set may include both base language (Java) instructions (e.g., method invocation instruction), and aspect-specific (AspectJ) instructions (e.g., advice execution instruction). The COMPUTE process takes an instruction $instr$ and produces a join point computation $x_{jp}$. The process also constructs a join point $desc_{jp}$ that describes the computation.

## 5. THE DUALITY OF OPEN CLASSES

Open Classes (OC) is an aspect mechanism found in AspectJ that allows aspects to change the structure of a concern program. The mechanism realizes the semantics of *inter-type declarations*. An inter-type declaration construct associates a target Java type (*what* to change) with an OC effect (*how* to change it).

There are two kinds of OC declaration effects: member and super type. A *member declaration* effect introduces a new member into the target Java type. For example, the following inter-type declaration introduces an observers field into the Point class:

```
public Vector Point.observers;
```

A *super type declaration* effect transforms the inheritance relationship of the target type. It is specified by a **declare parents** construct. For example, the following declaration resets Point's superclass to be Observable:

```
declare parents: Point extends Observable;
```

### 5.1 A Nonreactive OC Mechanism

We can explain the OC mechanism in terms of a nonreactive CRX weaving process (Figure 11). The concern program $P$ contains OC effects and base program types (including aspect classes); the weaving plan $\varrho$ maps target types to the effects; and $\wp$ is a Java program where the types and the effects are combined.

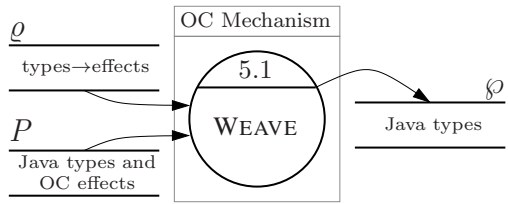

**Figure 11:** Nonreactive OC Mechanism

The WEAVE (5.1) process iterates over $P$ types. For each type, WEAVE selects the relevant OC effects using the $\varrho$ mappings. Then it transforms the type by applying the effects; and finally it commits the transformation result to the composed program.

### 5.2 A Reactive OC Mechanism

Interestingly, the OC mechanism can also be explained in terms of a *static* reactive CRX weaving process. A reactive OC mechanism iteratively constructs an AST of a composed program. For clarity, we focus on a simplified representation of an AST that has four types of nodes, namely (from parent to children), *root*, *type name*, *type declaration*, and *type member* (Figure 12).

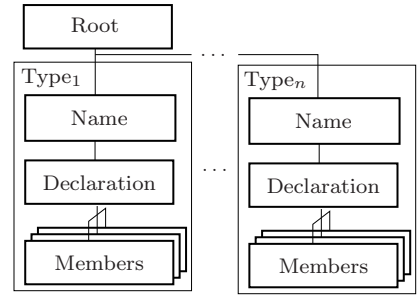

**Figure 12:** A simplified representation of a Java program's AST

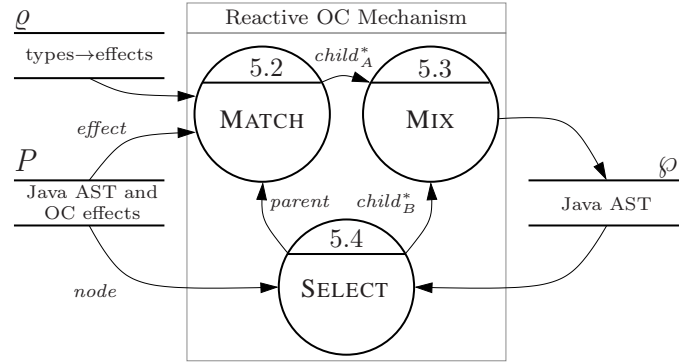

**Figure 13:** Reactive OC Mechanism

The DFD diagram in Figure 13 depicts an (alternative) reactive model for OC. The model combines three processes: MATCH (5.2), MIX (5.3), and SELECT (5.4). Initially, the composed program $\wp$ contains only a single root node marked as open. At each step, the mechanism selects an open *parent* node in $\wp$, populating its children with AST nodes drawn from $P$. That is, SELECT selects the *parent* from $\wp$ and the *parent*'s "base" child nodes $child^*_B$ from the AST of the concern program $P$. MATCH selects OC effects associated with *parent*, and translates them to $child^*_A$ nodes.[5] The MIX process integrates the base and the "advice" children together, and commits the result to the composed program. The child nodes are marked open, with the exception of type member nodes, which are marked closed. Weaving continues until all $\wp$ nodes are closed.

A reactive OC resembles PA. An open node *parent* corresponds to a join point $desc_{jp}$, $child^*_B$ correspond to a join point computation $x_{jp}$, and $child^*_A$ correspond to advice computations $x^*_{adv}$.

The nonreactive and reactive alternative OC models illustrate different *understandings* of this mechanism. OC neither changes type names nor adds or removes types. The concern program types always end up in the composed program under the same name. This allows one to interpret the OC weaving plan mappings in two ways: as a mapping between *concern* types and effects, or as a mapping between *composed* types and effects. The first interpretation implies a *nonreactive* weaving process; the second, a *reactive* one.

---
[5]$child^*_A$ are *parent*'s children defined by inter-type declarations. For example, Vector Point.observers defines the observers field as a child of the Point class declaration.

# 6. BEYOND CONCEPTUAL MODELS

The CRX model affords deeper understanding of an aspect mechanism and its internal weaving process.

## 6.1 Abstract Domains

In CRX, the abstract AOP language representation overcomes the syntax dependency problem found in ABX. CRX defines the mechanism in terms of concern elements (domain $\mathcal{C}$) and integration rules (domain $\mathcal{R}$). These domains abstract away from any concrete representation. A program in a concrete AOP language syntax, e.g., in AspectJ, can be preprocessed, e.g., by a READ (6.0) process (Figure 14) that reads $p_A$ and $p_B$ in and writes $P$ and $\varrho$ out, which in turn can be used by WEAVE (4.0, Figure 8). Consequently, the CRX weaving process model generalizes over asymmetric (e.g., AspectJ) and symmetric (e.g., Hyper/J, AspectWerkz, and Classpects) AOP languages uniformly.

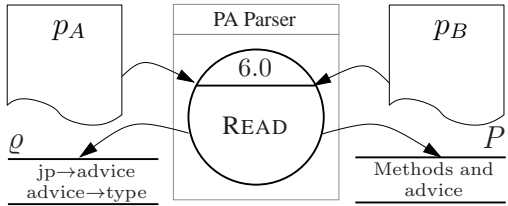

**Figure 14: Syntactical versus semantical**

## 6.2 WPM as a Unifying Concept

The CRX process model reveals that join points are only used in reactive mechanisms. Nonreactive mechanisms do not use join points because they do not read from the composed program register $\mathcal{X}$. In other words, the existence of a join point abstraction can be used to *distinguish* rather than *unify* various aspect mechanisms, e.g., PA is a join point-based mechanism, CMP is not.

A CRX weaving process model was constructed top-down using abstract terms. The abstract terminology is useful for comparing different weaving processes. Zooming out highlights the similarities. Zooming in exposes differences. For example, at the top level both CMP and PA realize the same CRX weaving process. However, they fall into two different categories. CMP is a nonreactive, whereas PA is a reactive mechanism.

## 6.3 Cyclic Dependency in AspectJ

AspectJ has a cyclic dependency between a computation in $X$ and the join point in $X_{JP}$ that describes that computation. The CRX weaving model (recall Figure 10) explains how the cyclic dependency is resolved in the PA weaving process. Specifically, the join point $desc_{jp}$ describes the join point computation $x_{jp}$ rather than the composed program computation. Thus, the join point is created before the composed computation is constructed.

## 6.4 Advising Advice Execution

ASB's view that PA selects and merges base and aspect program elements at each join point does not explain weaving at advice execution join points. Clearly, at advice execution join points, *no* base element should be selected. The role of base is played by a piece of *advised* advice. ASB does not generalize over advice execution join points because it explains *advised* elements strictly as *base*.

In contrast, the CRX weaving model of PA (Figure 10) distinguishes between a join point computation (an advised computation), and an advice computation. The join point computation ($x_{jp}$) is produced by the evaluator's COMPUTE process as a result of running $\wp$ computations. Since $\wp$ generally contains previously woven advice computations, the join point computation might be an advice execution computation. Thus, the CRX weaving process model also explains how advice executions are advised.

## 6.5 Untangling Evaluate and Advise

The reactive CRX model of PA explicitly separates EVALUATE (interpreting process) from ADVISE (advising process). EVALUATE selects and computes the meaning of advised elements (e.g., methods). ADVISE selects advice, computes their meaning, and weaves them. In contrast, ASB modifies the interpreter to include aspectual operations. For example, in ASB's PA mechanism, calls to `lookup-advice` are scattered throughout the interpreter and tangled with calls to `lookup-method`.

# 7. TOP-DOWN CLASSIFICATION

Our top-down analysis yields a process-oriented classification of aspect mechanisms. At the top level, we distinguish between a *reactive* and a *nonreactive* mechanism. In this section we compare our classification to existing categorizations of AOP languages. We show that our classification is different, and explain its usefulness.

## 7.1 Symmetric versus Asymmetric Languages

Today, existing AOP languages are sometimes categorized using the *syntactic* property of symmetry. Under this view, AspectJ is *asymmetric*—an AspectJ program consists of a base Java program and a set of aspect definitions. Hyper/J is said to be *symmetric*—all concerns are written in Java.

Unfortunately, this view does not provide much insight into the essential difference between AOP languages. It fails to explain how Hyper/J differs from AspectJ on a *semantical* level. Worse yet, under the symmetry-based view, "symmetrical" AspectJ-like languages (e.g., AspectWerkz and Classpects [17]) fall into the category of Hyper/J, instead of sharing a category with AspectJ.

Our approach abstracts away from the syntax of a language, and focuses instead on the weaving process that realizes its semantics. We represent an input AOP program in a syntax-independent manner, through sets of concern elements and integration rules.

We explain the working of an aspect mechanism according to its semantical operation. The reactive PA weaving process (Figure 10) explains AspectWerkz and Classpects in exactly the same manner as it explains AspectJ. Our classification identifies that AspectWerkz, Classpects, and AspectJ are all reactive aspect mechanisms, and contrasts them with the nonreactive Hyper/J CMP.

## 7.2 Static versus Dynamic Languages

Another classification approach categorizes AOP languages as *static* or *dynamic*. A static language expresses a weaving process over an abstract syntax tree (AST). For example, Hyper/J dictates the construction of a hypermodule unit tree from a hyperspace unit tree, where units represent AST nodes. Hyper/J is often referred to as a *static* AOP language. A dynamic AOP language allows the expression of concern integration logic over computations, using run-time abstractions. For example, advice weaving in AspectJ is defined in terms of join points in the program execution. AspectJ is often said to be a *dynamic* AOP language.

In our approach the weaving problem and the general CRX weaving process model are abstracted in terms of concern and composed elements. Identifying the composed and concern elements as AST nodes yields a weaving model for static languages. Defining the elements to be computations yields a weaving model for dynamic languages. Although it is tempting to think that nonreactive aspect mechanisms are always static, and reactive aspect mechanisms are always dynamic, it is not so. Specifically, the *static* OC mechanism can realize a *reactive* weaving process. A reactive–nonreactive classification is, therefore, fundamentally different than a static–dynamic one.

### 7.3 Static versus Dynamic Semantics

Aside from the static–dynamic classification of AOP languages, there is an analogous classification of AOP language semantics. Semantics for dynamic AOP language can be specified in terms of code transformation. It is also possible for static languages or mechanisms to be given dynamic semantics.

In our approach, a weaving process always maps to a specific semantics, either static or dynamic. Alternative semantics for the same AOP language are normally realized by different weaving processes. For example, compilation semantics for AspectJ can be realized by a static nonreactive mechanism; dynamic semantics for AspectJ can be realized by a dynamic reactive mechanism.

## 8. TOP-DOWN CONSTRUCTION

A property of our top-down approach for modeling a weaving process is the ability to systematically construct new aspect mechanisms. The mechanisms can be derived in a step-wise refinement of the general CRX weaving process model (Figure 5). For example, the reactive PA mechanism model (Figure 10), which refines the general CRX weaving process model, can be further specialized to produce a concrete weaving process specification.

### 8.1 Component-Based Design

In our approach an aspect mechanism is modeled by a set of collaborating subprocesses. A component-based design [12] for an aspect mechanism would support a decomposition of an aspect mechanisms into distinct processes. In such a component-based design, it would be possible to replace and reuse subprocesses, thus facilitating aspect mechanism evolution. For example, the reactive PA weaving process model decouples the interpreter functionality from the aspectual advice selection and weaving processes [13]. The interpreter operations are realized by CONTROL (4.2.2) and COMPUTE (4.2.1). MATCH (4.1.1) and MIX (4.1.2) realize the match and mix aspectual functionality.

Decomposing the aspect mechanism into subcomponents affords replacement of one component at a time. Consider two reasonable enhancements to AspectJ: extending the pointcut designators with new regular expressions, and adding new types of advice. Each enhancement requires replacement of exactly one component. Extending the original MATCH component to handle new regular expressions would enable the pointcut designators enhancement. To extend the advice type set, the MIX component should be upgraded.

### 8.2 New AOP Functionality

Most (if not all) the weaving processes that exist to date construct the composed program by extending it with new elements. For example, the advice weaving process in PA simply extends, at each step, the composed program's computation trace with new computations. AspectJ never transforms computations within the composed program (i.e., an already executed or currently active computations). Similarly, the weaving processes of CMP and OC never change elements of the composed program.

The general CRX weaving process model (Figure 5) lends itself toward inventing novel weaving functionality beyond current practices. The model defines an aspect mechanism as a composed program transformer. At each step of the weaving process, the mechanism transforms the composed program $\wp$. Under this view, the process may include steps that actually modify or replace elements in $\wp$. For example, one can imagine a dynamic reactive process that would update currently active computations (e.g., transform the current continuation). Another example would be a static reactive process that can extend $\wp$ with new AST nodes, and transform previously composed AST nodes.

## 9. CONCLUSION

This paper characterizes the design space for aspect mechanisms. An aspect mechanism is an implementation of an aspect weaver. We model weaving as the process of concern integration, and define the weaving process in an abstract way, independent of any specific aspect mechanism. The abstract model describes the design space, and concrete aspect mechanisms are derived top-down from the abstract model. We distinguish between reactive and non-reactive mechanisms. This taxonomy provides a common framework for comparing and contrasting different aspect mechanisms.

We offer a new CRX model as the next evolutionary step in modeling aspect mechanisms. In CRX, integration rules in $\mathbf{R}$ govern the process of integrating concerns in $\mathbf{C}$ into crosscutting structures in $\mathbf{X}$. We formalize the CRX model as a weaving problem over corresponding abstract domains $\mathcal{R}$, $\mathcal{C}$, and $\mathcal{X}$. A CRX weaving process executes a solution to a weaving problem.

The weaving processes are classified as either reactive or nonreactive, instead of the more classic static–dynamic or symmetric–asymmetric classifications of AOP languages. We design a nonreactive CRX process for the static CMP aspect mechanism of the symmetric Hyper/J language. We design a reactive CRX process for the dynamic PA aspect mechanism of the asymmetric AspectJ language. We also design both a nonreactive and a reactive CRX processes for the static OC aspect mechanism of AspectJ.

Our top-down model overcomes many limitations in other models in terms of abstraction, precision, and uniformity. The model was not generalized from a particular set of aspect mechanisms, yet existing mechanisms can be mapped, analyzed, and explained in terms of this model. In contrast, other models are generally constructed bottom-up by analyzing existing aspect mechanisms. The generality of these bottom-up models is illustrated by expressing in terms of the model only the mechanisms used to derive them.

The design space for CRX aspect mechanisms can help not only classify existing mechanisms but also develop new ones. It can guide the implementor of new aspect mechanisms in existing classes. It can also guide the designer when mechanisms implementing new kinds of weaving are needed. We have used the CRX model in researching third-party composition of aspect mechanisms. The model was particularly useful in resolving sequential black-box composition of heterogeneous aspect mechanisms [7] and parallel glass-box composition of homogeneous aspect mechanisms [11]. Future work may include extending the CRX model to also describe a multi-mechanism weaving process.

We believe that the CRX model is a good way to explain and teach AOP [10]. Data flow analysis is effective for conceptual level modeling, and a DFD provides an easy to understand graphical representation of a system [24]. Of course, a data flow description of the weaving process is just a step toward a complete definition or a blueprint of an aspect mechanism. Documenting the weaving process also in terms of state transitions is another logical follow-up to this work.

### Acknowledgment

We are grateful to Kevin Sullivan for his helpful suggestions on this paper. We also thank Gene Cooperman, Westley Weimer, and the anonymous reviewers for their insightful comments.